\documentclass{PoS}
\newcommand{\be}{\begin{equation}}
\newcommand{\ee}{\end{equation}}

\def\be{\begin{equation}}
\def\ee{\end{equation}}
\def\beq{\begin{eqnarray}}
\def\eeq{\end{eqnarray}}

\def\({\left (}
\def\){\right )}
\def\[{\left [}
\def\[{\right ]}

\def\ba{\begin{eqnarray}}
\def\ea{\end{eqnarray}}

\title{Impact of coherence and low x effects on decorelations of forward-central jets.}

\ShortTitle{Coherence and low $x$}

\author{\speaker{Krzysztof Kutak}\\
       Institute for Nuclear Physics, Polish Academy of Science\\
Radzikowskiego 152, Krakow, Polska.\\
        E-mail: \email{krzysztof.kutak@ifj.edu.pl}}

\abstract{We report on recent calculation of forward central jets decorelations at LHC energies within High Energy Factorization. We emphasize the role of  Sudakov resummation and kinematical effects in description of data.}

\FullConference{XXII. International Workshop on Deep-Inelastic Scattering and Related Subjects,\\
		28 April - 2 May 2014\\
		Warsaw, Poland}

\begin{document}

\section{Introduction}

Here we report on recent studies \cite{vanHameren:2014ala}, of decorelations in production of central-forward dijet the High
Energy Factorization (HEF) approach.
This observable has been recognized to be very usefull in  analysis of parton dynamics at small-$x$ and not negligible transversal momentum of initial state gluons~\cite{Vera:2007kn,Marquet:2003dm,Deak:2009xt,Deak:2010gk,Kutak:2012rf,Nefedov:2013ywa,vanHameren:2014lna}.
One of the reasons is that small-$x$ effects are inseparably related to the notion
of internal transverse momenta of gluons (which due to that are off-shell) inside a hadron, which, according to
the Balitsky-Fadin-Kuraev-Lipatov (BFKL) formalism, can be
large but cannot be arbitrarily small because of the importance of nonlinear effects
absent in the BFKL equation and taken in to account by the BK equation. 
Furthermore the internal
transverse momentum of a gluon can be viewed as a direct source of azimuthal
decorrelations, since it creates a jet momentum imbalance on the transverse plane.

 It seems the effects that turn out to
play a crucial role in the description of data are the kinematical effects enforcing momenta of the gluons
to be dominated by the transverse
components~\cite{Kwiecinski:1996td,Andersson:1995jt} and the coherence effects effects related to
soft gluon radiation of Sudakov type \cite{Sudakov:1954sw}.

In situations where the final state populates forward rapidity regions, one of
the longitudinal fractions of the hadron momenta is much smaller then the other,
$x_{A}\ll x_{B}$, and the following `hybrid' HEF formula is
used~\cite{Deak:2009xt}
\be
d\sigma_{AB\rightarrow X}=\int \frac{d^{2}k_{TA}}{\pi}\int\frac{dx_{A}}{x_{A}}\,\int dx_{B}\, \\
\sum_{b}\mathcal{F}_{g^{*}/A}\left(x_{A},k_{TA},\mu \right)\, f_{b/B}\left(x_{B},\mu \right)\,
 d\hat{\sigma}_{g^{*}b\rightarrow X}\left(x_{A},x_{B},k_{TA},\mu \right),\label{eq:HENfact_2}
\ee
where $\mathcal{F}_{g^{*}/A}$ is a UGD (Unintegrated Gluon Density), $f_{b/B}$ are the collinear PDFs, and
$b$ runs over the gluon and all the quarks that can contribute to the production
of a multi-particle state $X$. 
Note that both $f_{b/B}$ and $\mathcal{F}_{g^{*}/A}$ depend on the hard
scale $\mu$.
 As we explain below, it is important to incorporate the hard scale dependence also in UGD.
The off-shell gauge-invariant matrix elements for multiple final states reside in $d\hat{\sigma}_{g^{*}b\rightarrow X}$. 
The condition $x_{B}\gg x_{A}$ is imposed by
proper cuts on the phase space of $X$.
In our computations, we used several different unintegrated gluon densities
$\mathcal{F}_{g^{*}/A} (x, k_{T},\mu)$:
\begin{itemize}
  \item
  The nonlinear KS (Kutak-Sapeta) unintegrated gluon
  density~\cite{Kutak:2012rf}, which comes from the extension of the BK
  (Balitsky-Kovchegov) equation \cite{Kutak:2003bd} following the prescription
  of KMS (Kwiecinski-Martin-Stasto) 
  \item
  The linear KS gluon~, determined from the linearized
  version of the equation described above.
  \item
  The KMR hard scale dependent unintegrated gluon density~\cite{Kimber:1999xc}.
  It is obtained from the standard,
  collinear PDFs supplemented by the Sudakov form factor 
  and small-$x$ resummation of the BFKL type.
 The Sudakov form factor ensures no
  emissions between the scale of the gluon transverse momentum, $k_T$, and
  the scale of the hard process, $\mu$. 
  \item
  The standard collinear distribution $\mathcal{F}_{g^{*}} (x, k_{T}, \mu^2) =
  xg(x,\mu^2)\delta(k_T^2)$, which, when used in Eq.~(\ref{eq:HENfact_2}),
  reduces it to the collinear factorization formula. In this study we used the
  CTEQ10 NLO PDF set~\cite{Lai:2010vv}.

\end{itemize}

In addition, we supplement the KS linear and nonlinear UGDs with the Sudakov
resummation, which, as we shall see, turns out to be a necessary ingredient
needed to describe the data at 
moderate $\Delta\phi$.  The resummation is made on top of the Monte Carlo
generated events and it is motivated by the KMR prescription of the Sudakov form
factor. It effectively incorporates the dependence on a hard scale $\mu$ into the KS gluons, which by themselves do not exhibit such dependence.

\section*{Results}

\begin{figure}[t]
  \begin{center}
    \includegraphics[width=0.45\textwidth]{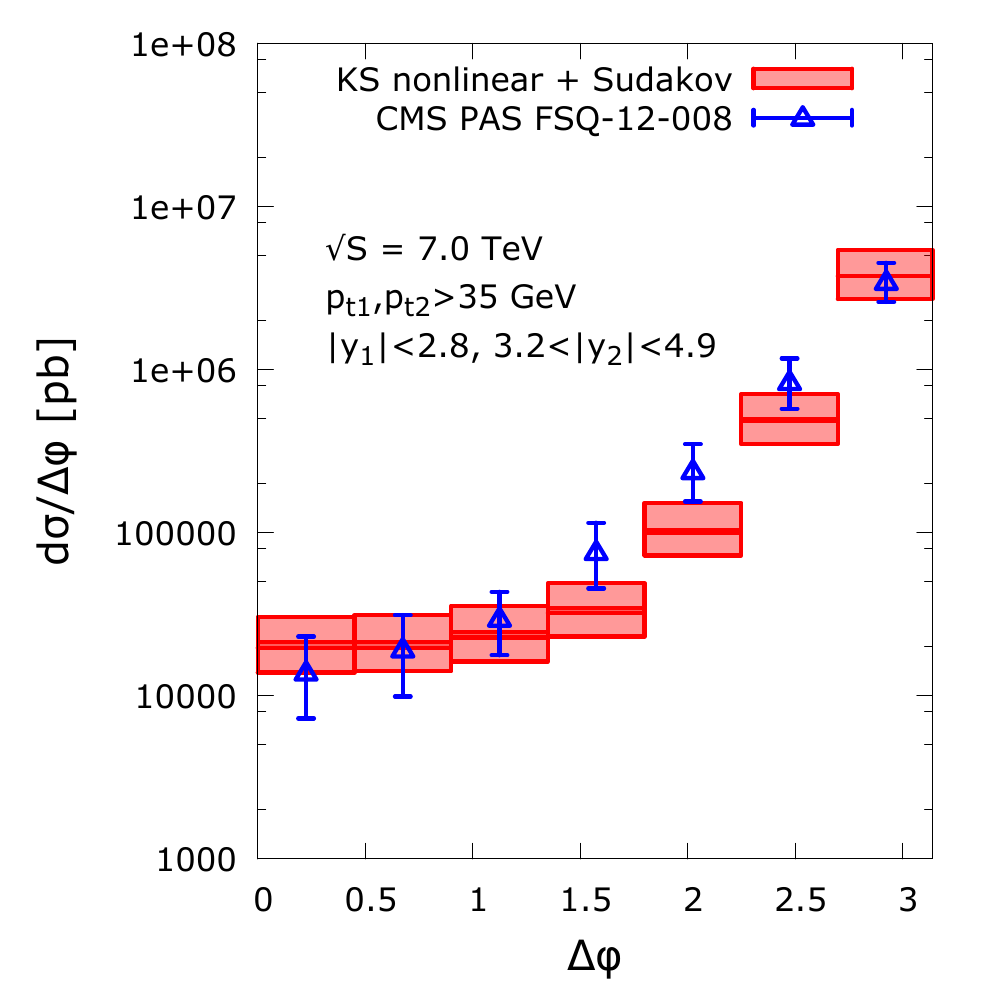}
    \includegraphics[width=0.45\textwidth]{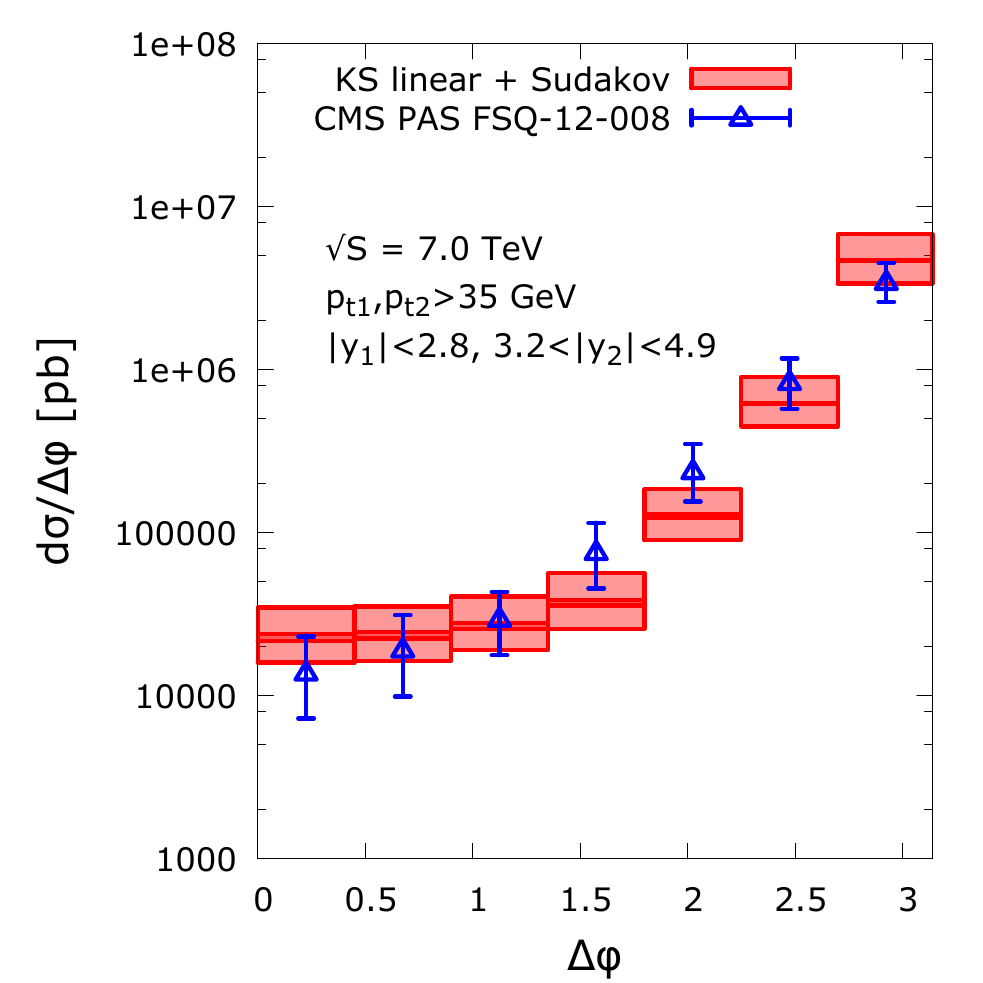} \\
    \includegraphics[width=0.45\textwidth]{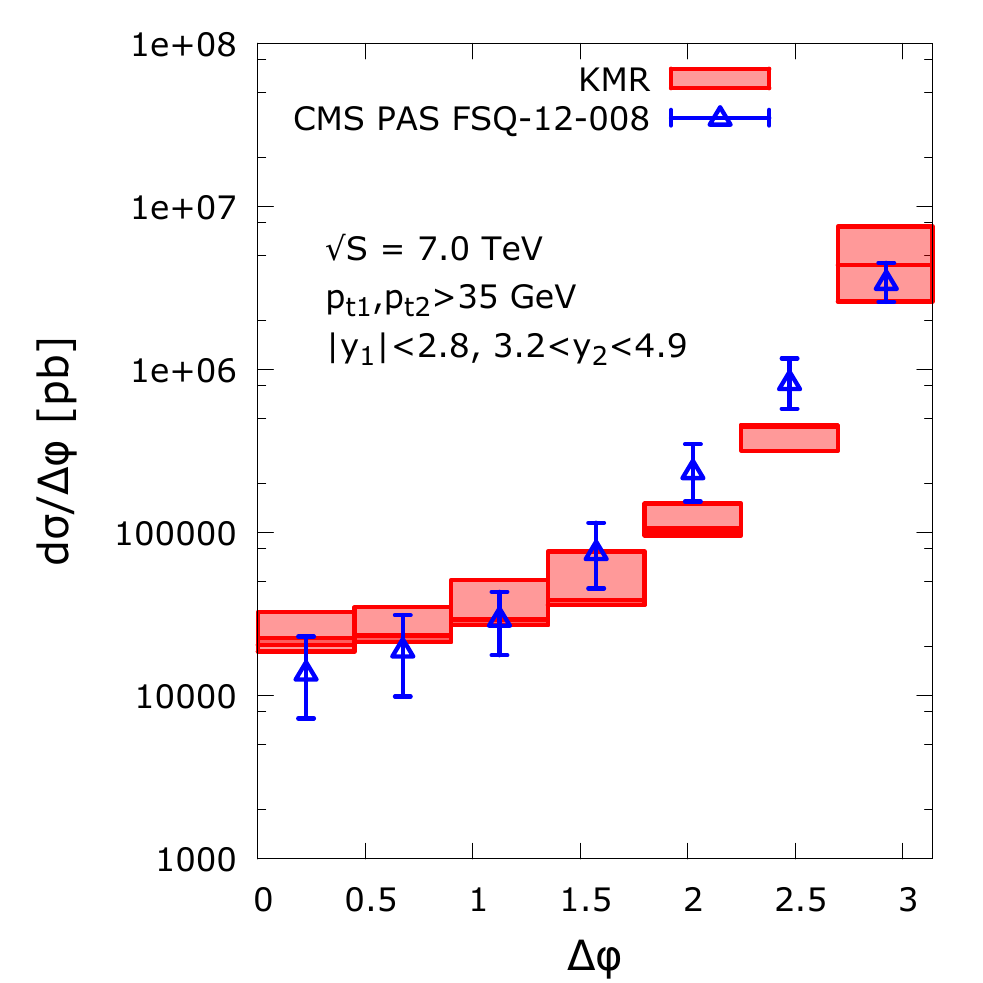} 
    \includegraphics[width=0.45\textwidth]{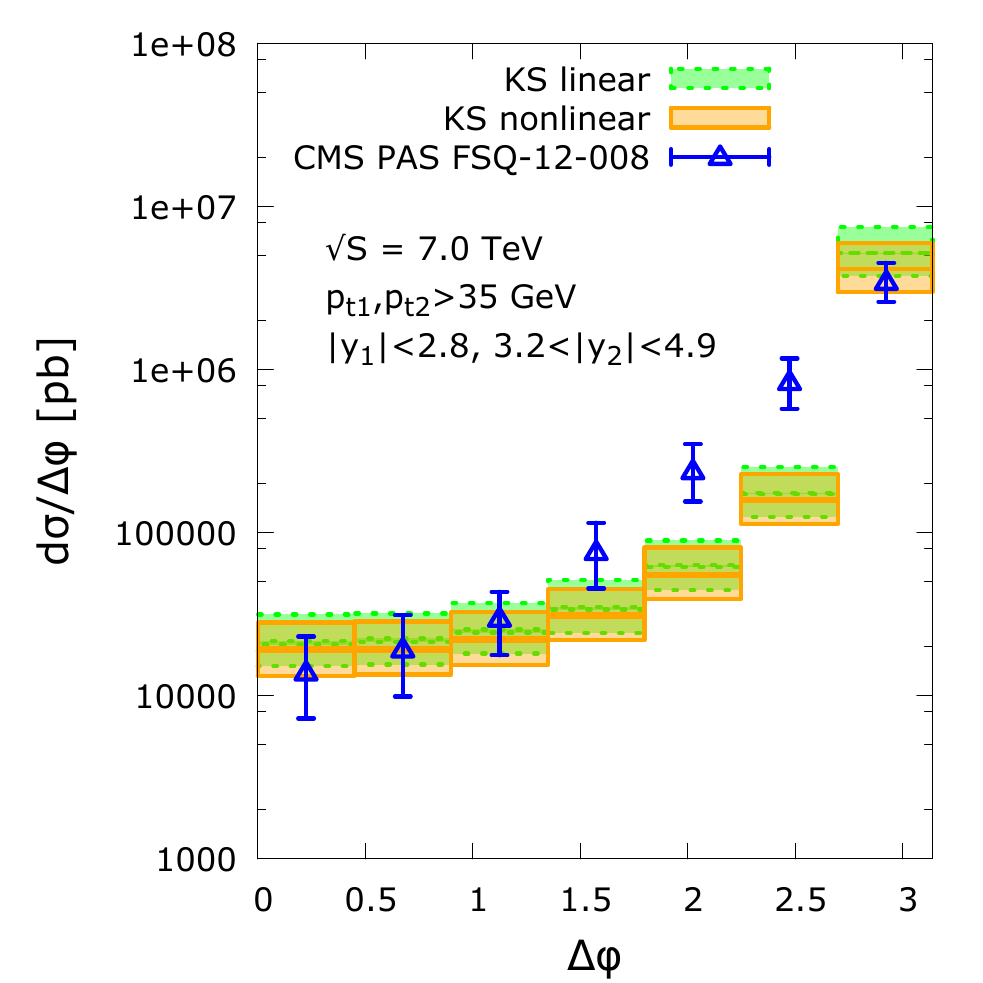} 
  \end{center}
  \caption{
  \small
  Comparison of CMS data for the inclusive dijet scenario with
  model predictions. 
  }
  \label{fig:veto_models_vs_CMS}
\end{figure}

\begin{figure}[t]

  \begin{center}
    \includegraphics[width=0.45\textwidth]{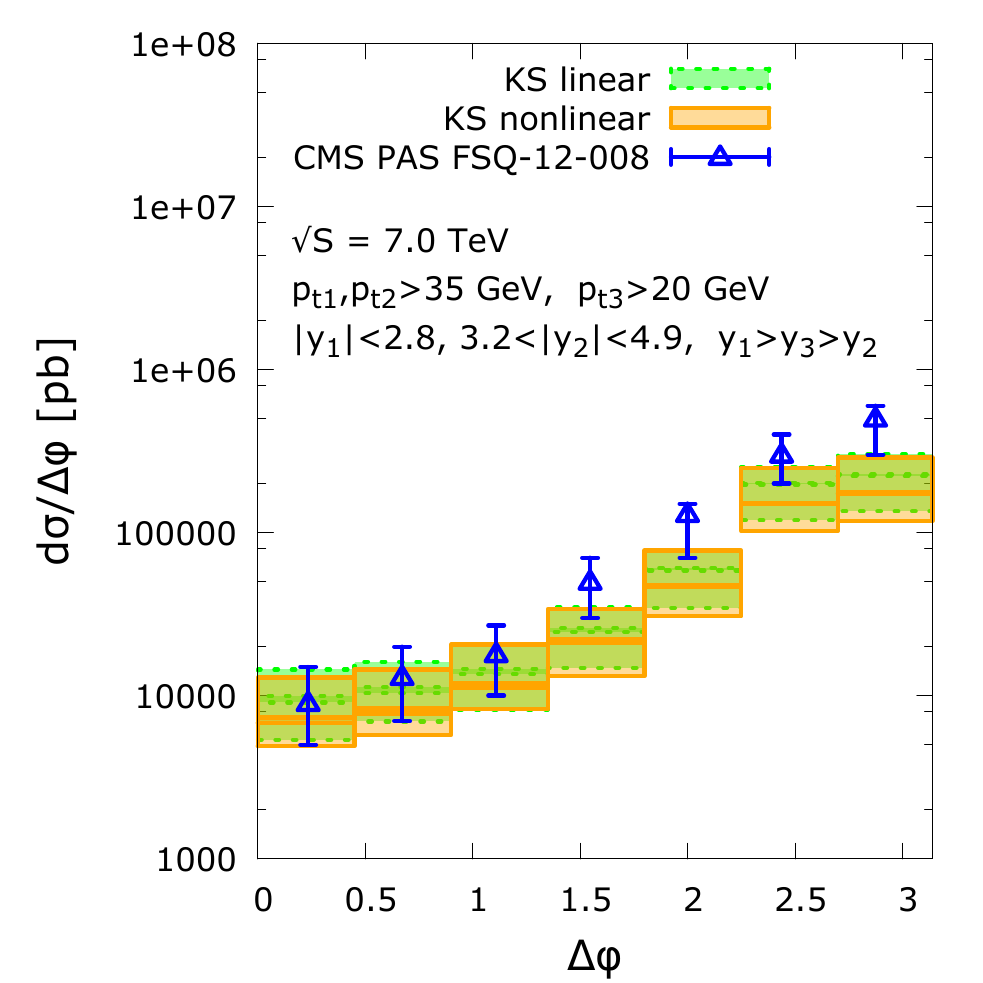}
    \includegraphics[width=0.45\textwidth]{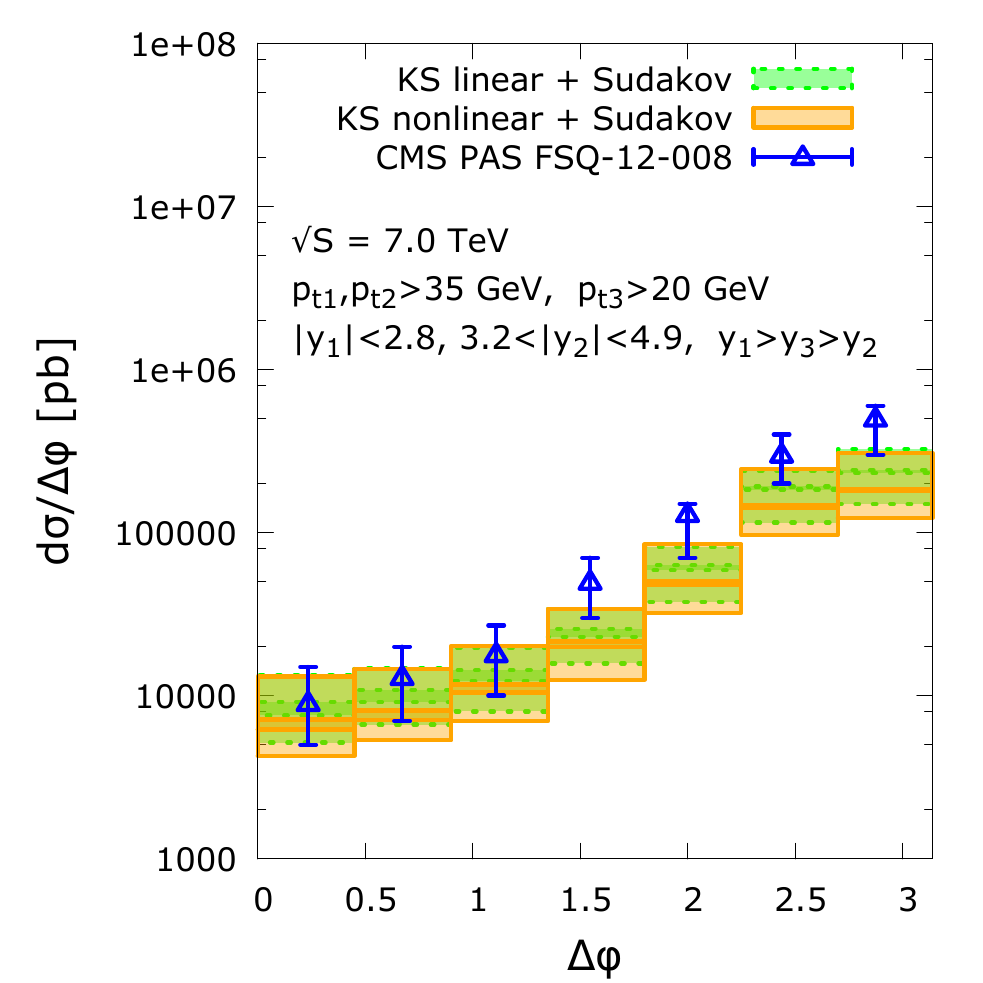} \\
    \includegraphics[width=0.45\textwidth]{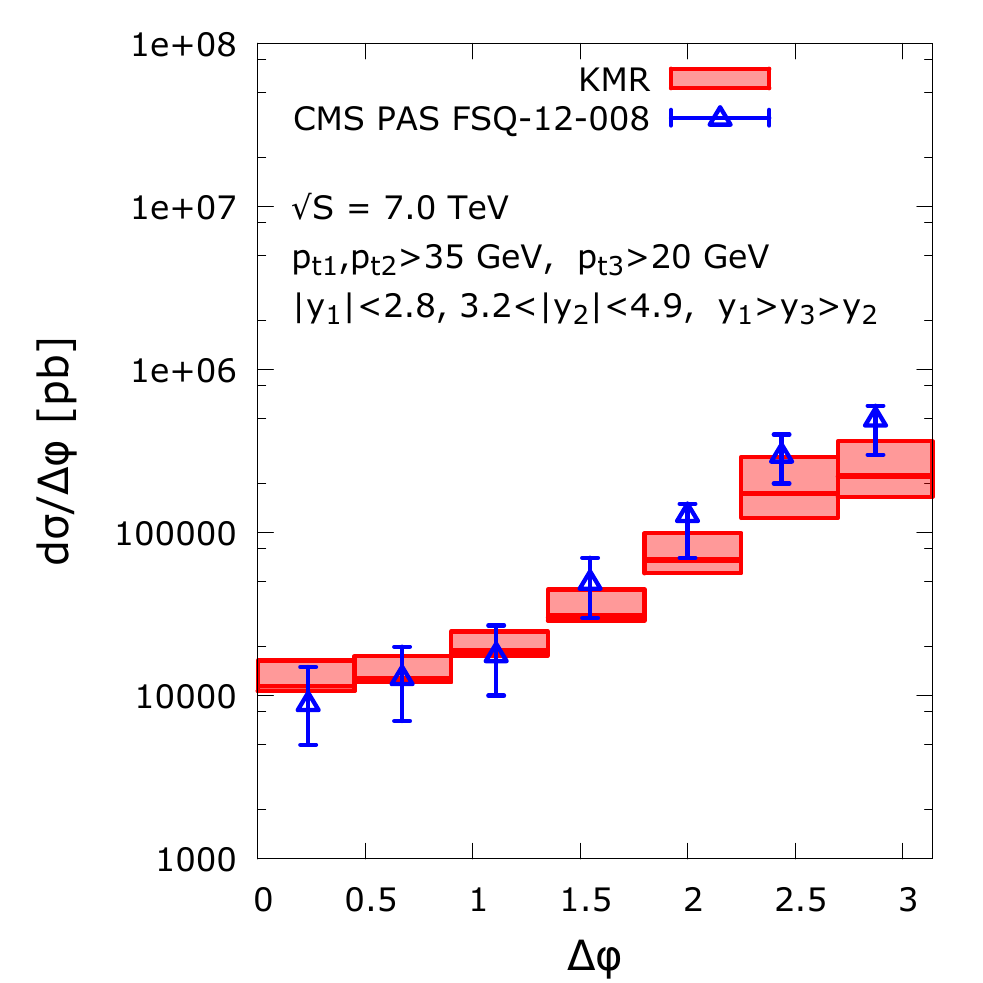} 
    \includegraphics[width=0.45\textwidth]{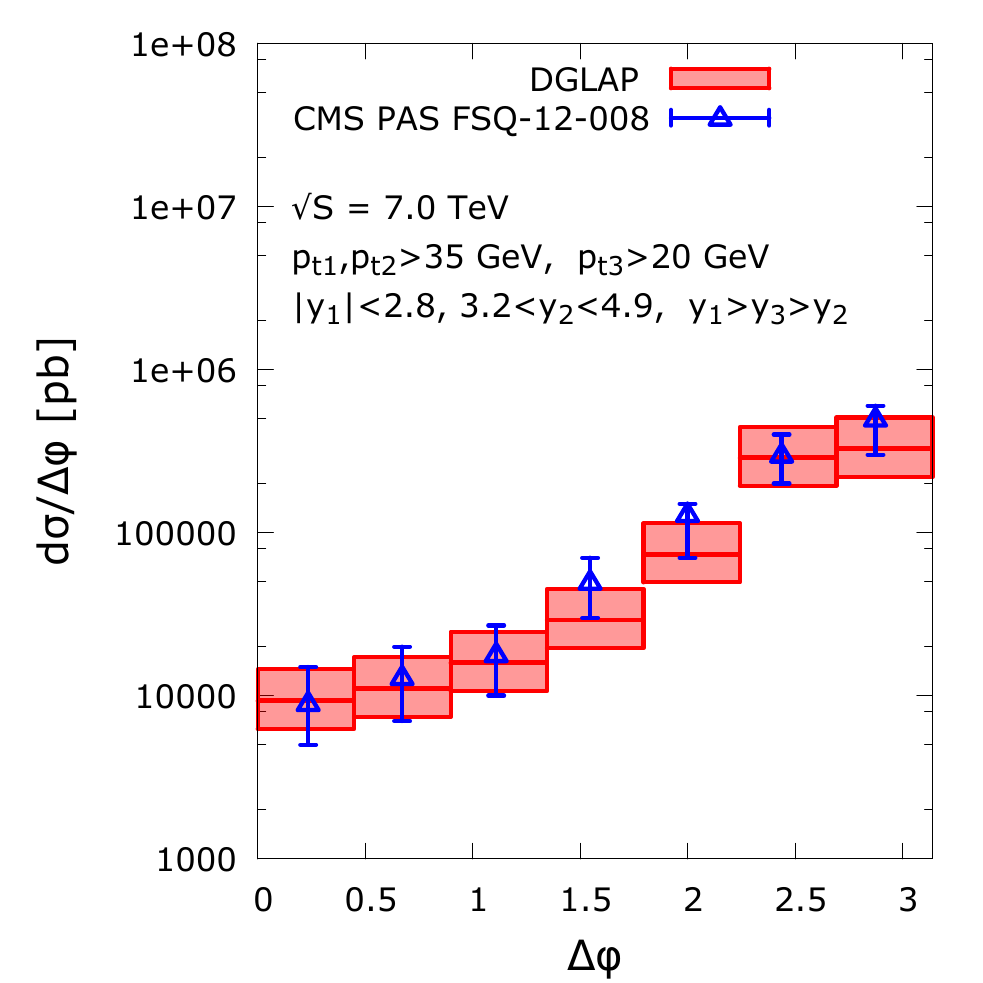} 
  \end{center}
  \caption{
  \small
  Comparison of CMS data for the dijet inside-jet-tag scenario with model 
  predictions. 
  }
  \label{fig:tag_models_vs_CMS}
\end{figure}

In this section, we present the results of our study of the azimuthal
decorrelations in the forward-central dijet production. Our framework enables
us to describe two scenarios considered in the  CMS forward-central dijet
measurement \cite{CMS:2014oma}: 
\begin{itemize}
\item 
\emph{Inclusive scenario}, which, in the experiment, corresponds to
selecting events with the two leading jets satisfying the cuts:  $p_{T,1,2} >
35\,$ GeV, $|y_1| < 2.8$, $3.2<|y_2|<4.9$ and with no extra requirement on 
further jets. These results are shown in Fig.~\ref{fig:veto_models_vs_CMS}.
\item
\emph{Inside-jet-tag scenario}, with the same selection on the two hardest jets
but, this time, a third jet with $p_T>20$ GeV is required between the forward
and the central region. The corresponding results are shown in
Fig.~\ref{fig:tag_models_vs_CMS}.
\end{itemize}
 
In Fig.~\ref{fig:veto_models_vs_CMS}, we present our 
results
 for the case of
the inclusive selection and compare them with the data from CMS. We show the
results obtained with the nonlinear and linear KS gluon, supplemented with the
Sudakov form factor (top left and top right, respectively), the KMR gluon
(bottom left) and an unmodified KS gluon (bottom right). We see that the
KS+Sudakov and KMR describe the data well. The error bands on the predictions
were obtained by varying the hard scale appearing explicitly in the running
coupling, UGDs, and the collinear PDFs by a factor $2^{\pm 1}$.
The calculations were performed independently by three programs:
$\mathtt{LxJet}$ \cite{Kotko_LxJet},  $\mathtt{forward}$~\cite{forward}, and a
program implementing the method of \cite{vanHameren:2012if}.

The results presented in Fig.~\ref{fig:veto_models_vs_CMS} provide evidence
in favour of the small-$x$ (or BFKL-like) dynamics. This dynamics produces gluon
emissions, unordered in $k_T$, which build up the non-vanishing $\Delta\phi$
distributions away from $\Delta\phi=\pi$. (A pure DGLAP based approach, without the
use of a parton shower, could of course only produce a delta function at
$\Delta\phi=\pi$.)
Furthermore, combining the above result with the recent analysis performed in
\cite{vanHameren:2014lna}, we conclude that the effects of higher
orders, like kinematical effects that allow for emissions at low $\Delta\phi$
are of crucial importance.
This alone is however not enough, since, as shown in
Fig.~\ref{fig:veto_models_vs_CMS}, one necessarily needs the Sudakov resummation
to improve the moderate $\Delta\phi$ (or equivalently moderate $k_T\sim 50\,
$ GeV) region.
These Sudakov effects are needed to resum virtual emissions between the hard
scale provided by the external probe and the scale of the emission from the gluonic
ladder. In other words, one has to assure that the external scale is the largest
scale in the scattering event. 

Fig.~\ref{fig:tag_models_vs_CMS} shows the results obtained in the HEF approach
with the $2\rightarrow 3$ hard matrix elements. We also show the corresponding
result from pure DGLAP, i.e.\ with the HEF formula (\ref{eq:HENfact_2}) used in
the collinear limit.  We see that the linear and nonlinear KS results without
(top left) and with (top right) the Sudakov form factor nicely follow the experimental data
from the inside-jet-tag scenario. The description is also very good when the KMR
gluon is used (bottom left).  In the case of pure 
DGLAP calculation
(bottom right), the
parton produced in the final state allows for generation of the
necessary transverse momentum imbalance between the two leading jets. This, in
turn, leads to a good description of the experimental data, even without the use
of a parton shower.\\

Our results confirm the necessity of incorporating the hard scale dependence
to the unintegrated gluon densities. 
Further studies with in CCFM-based approaches~\cite{Jung:2010si,Hautmann:2013tba}
would be needed in order to get better understanding of these effects.

\section*{Acknowledgments}
The work presented in DIS 2014 is based on resarch results obtained in collaboration with Anderas van Hameren, Piotr Kotko,
Sebastian Sapeta.\\
This research has been supportrd by NCBiR Grant No. LIDER/02/35/L-2/10/NCBiR/2011.


\begin{thebibliography}{99}
\bibitem{forward} S. Sapeta. FORWARD, the code is available on request from the author
\end{thebibliography}
\end{document}